\RequirePackage{fix-cm}
\documentclass[twocolumn,epjc3]{svjour3}  
\smartqed  
\RequirePackage{graphicx}
\usepackage{amsmath}
\usepackage{dsfont}
\usepackage{enumerate}
\usepackage{lineno}

\usepackage[%
  verbose,
  colorlinks=true,
  naturalnames=true,
  linkcolor=blue,
]{hyperref}
\journalname{Eur. Phys. J. C}
\begin{document}

\title{Generalisation of the identity method for determination of high-order moments of multiplicity distributions with a software implementation
}


\author{Maja Ma\'{c}kowiak-Paw{\l}owska\thanksref{e1}
        \and
        Piotr Przyby{\l}a\thanksref{addr2} 
}

\thankstext{e1}{e-mail: majam@if.pw.edu.pl}


\institute{Faculty of Physics, Warsaw University of Technology,
 Warsaw, Poland\label{addr1}
           \and
           School of Computer Science, University of Manchester, Manchester, UK \label{addr2}
}

\date{Received: date / Accepted: date}

\maketitle

\begin{abstract}
The incomplete particle identification limits the experimentally-available phase space region for identified particle analysis. This problem affects ongoing fluctuation and correlation studies including the search for the critical point of strongly interacting matter performed on SPS and RHIC accelerators. In this paper we provide a procedure to obtain $n$th order moments of the multiplicity distribution using the identity method, generalising previously published solutions for $n=2$ and $n=3$. Moreover, we present an open source software implementation of this computation, called \textit{Idhim}, that allows one to obtain the true moments of identified particle multiplicity distributions from the measured ones provided the response function of the detector is known.
\end{abstract}

\section{Introduction}
Search for the critical point of strongly interacting matter remains one of the most important goals of experimental searches in heavy ion physics~\cite{Antoniou:2006mh,STARBESIIWP}. Its basic property --- the increase of the correlation length of the considered system --- forces experimenters to shift their interests from inclusive spectra to higher-order moments and cumulants of the particle multiplicity distributions. A particular interest is paid towards net-proton fluctuations being the most sensitive to the searched phenomenon~\cite{Hatta:2003wn}. 

One of the most serious experimental issues, which largely limits the available phase-space coverage, and, possibly, affects the studied signal, is the incomplete particle identification caused by finite detector resolution. To overcome this problem an experimental technique, called {\it the identity method}, was proposed in Ref.~\cite{Gazdzicki:2011xz}, and extended in Refs.~\cite{Gorenstein:2011hr,Rustamov:2012bx,Pruneau:2017fim}. { So far the identity method was described for the second~\cite{Gorenstein:2011hr} and third~\cite{Rustamov:2012bx} order moments. In Ref.~\cite{Rustamov:2012bx} it was also used to reexamine the first moments of the identified particle distributions. The impact of particle losses due to detector inefficiencies on results from the identity method is discussed in Ref.~\cite{Pruneau:2017fim}. The author shows that it remains applicable provided detection efficiencies can be determined with sufficient accuracy. With the ongoing development of theoretical studies concerning higher order moments it seems appropriate to extend experimental techniques and tools as well.}

In the present study the identity method is extended in two ways. Firstly, a strict procedure to obtain $n$th order moments of the multiplicity distribution is shown. 
Secondly, a program, called {\it Idhim}, which performs such calculations for any given number of considered particle types, is presented. It also allows one to obtain moments up to any order provided the detector response function is known. The modification of first moments from Ref.~\cite{Rustamov:2012bx}, also included in \textit{Idhim}, may address possible biases in other popular methods (e.g. maximal likelihood method~\cite{Marco_fit2,vanLeeuwen:2003ru}). 

The paper is organized as follows. In Sect.~\ref{sec_basic}, basic quantities of the identity method are presented. The computation of $n$th moments of true multiplicity distribution is shown in Sect.~\ref{sec_computation}. Modifications necessary to apply the general formulas in practice are addressed in Sect.~\ref{sec_modifications}. Description of the \textit{Idhim} program which computes moments of the true multiplicity distribution is given in Sect.~\ref{sec_idhim}. Section~\ref{sec_tests} contains tests of the program with the detector response close to the ones measured in real experiments. Conclusion in Sect.~\ref{sec_conclusion} ends the paper. 

\section{Basic quantities}
\label{sec_basic}
The identity method is developed under the assumption that particles are identified by measuring quantity $x$ (e.g., a mass) of observed particles. Due to the finite detector resolution one gets a continuous distribution for $x$, denoted by $\rho_{j}(x)$, where index $j$ stands for one of $k$ particle types. The density is expected to sum up to the mean of $N_{j}$, i.e., the multiplicity for this type:
\begin{equation}
\label{eq:density}
\int \mathrm{d}x \rho_j(x)=\langle N_j \rangle,
\end{equation}
For a given particle observation, its conditional probability of being of a given type is expressed by a quantity called {\it identity}, defined as:
\begin{equation}
\label{eq:w}
w_j(x)\equiv\frac{\rho_j(x)}{\sum_{l=1}^k \rho_l(x)}.
\end{equation}
In the case of complete particle identification $w_{j}$ is reduced to two extreme values: $w_{j}=0$ for particles of types other than $j$ and $w_{j}=1$ for particles of type $j$.

In the same way, one can define an aggregated quantity for a given particle type:
\begin{equation}
W_j\equiv\sum_{i=1}^{N(\nu)}w_j(x_i),
\end{equation}
where $N(\nu)$ is the total multiplicity (including all particle types) of the $\nu$th of considered $N_{ev}$ events. From these events one obtains the distribution of different types of $W$ with its moments defined as
\begin{equation}
\langle W_1^{n_1}\cdot W_2^{n_2}\cdot\ldots\cdot W_k^{n_k} \rangle=\frac{1}{N_{ev}}\sum_{i=1}^{N_{ev}}W_1^{n_1}\cdot W_2^{n_2}\cdot\ldots\cdot W_k^{n_k}, 
\end{equation}
where $n_{j}$ denotes the order of the moment of the distribution of $W_{j}$.

\section{Computing the n-th moments of multiplicity distribution}
\label{sec_computation}
We will now show how one can compute all the $n$th moments of the multiplicity distribution $\langle N_1^{n_1}\cdot N_2^{n_2}\cdot\ldots\cdot N_k^{n_k} \rangle$ with $n_1+n_2+\dots+n_k=n$ using the moments of the measured  identity variables. The procedure will be a generalisation of those published for $n=2$ \cite{Gorenstein:2011hr} and $n=3$ \cite{Rustamov:2012bx}. First, we shall demonstrate how the value of a moment of identity variables $\langle W_1^{n_1}\cdot W_2^{n_2}\cdot\ldots\cdot W_k^{n_k} \rangle$, depends on the multiplicity distribution. We have the following:
\begin{multline}
\label{e1}
\langle W_1^{n_1}\cdot W_2^{n_2}\cdot\ldots\cdot W_k^{n_k} \rangle=\sum_{N_1=0}^\infty\sum_{N_2=0}^\infty\dots\sum_{N_k=0}^\infty\\
\mathrm{P}(N_1,N_2,\dots, N_k) \int \mathrm{d}x_1^1\mathrm{P}_1(x_1^1)\dots\int \mathrm{d}x_{N_1}^1\mathrm{P}_1(x_{N_1}^1)\\
\int \mathrm{d}x_1^2\mathrm{P}_2(x_1^2)\dots\int \mathrm{d}x_{N_2}^2\mathrm{P}_2(x_{N_2}^2)\cdots\\
\int \mathrm{d}x_1^k\mathrm{P}_k(x_1^k)\dots\int \mathrm{d}x_{N_k}^k\mathrm{P}_k(x_{N_k}^k)\\
\big[w_1(x_1^1)+\ldots+w_1(x_{N_1}^1)+w_1(x_1^2)+\ldots+w_1(x_{N_2}^2)\\
+\cdots+w_1(x_1^k)+\ldots+w_1(x_{N_k}^k)\big]^{n_1}\times\\
\big[w_2(x_1^1)+\ldots+w_2(x_{N_1}^1)+w_2(x_1^2)+\ldots+w_2(x_{N_2}^2)\\
+\cdots+w_2(x_1^k)+\ldots+w_2(x_{N_k}^k)\big]^{n_2}\times\cdots\times\\
\big[w_k(x_1^1)+\ldots+w_k(x_{N_1}^1)+w_k(x_1^2)+\ldots+w_k(x_{N_2}^2)\\
+\cdots+w_k(x_1^k)+\ldots+w_k(x_{N_k}^k)\big]^{n_k},
\end{multline}
where $\mathrm{P}(N_1,N_2,\dots, N_k)$ is the multiplicity distribution, i.e., the probability of observing $N_1$ particles of the first time, $N_2$ particles of the second type and so forth, and $\mathrm{P}_j(x)=\frac{\rho_j(x)}{\langle N_j\rangle}$ is the probability distribution of the $j$th type.

Let us firstly focus on the innermost part of equation \ref{e1}, denoted hereafter by $\omega$:
\begin{equation}
\omega\equiv\prod_{l=1}^{k}\left[\sum_{j=1}^{k}\sum_{i=1}^{N_j}w_l\left(x_i^j\right)\right]^{n_l}.
\end{equation}
We will now use the multinomial theorem to expand the $n_l$th power. Let us first define the following notation for brevity:
\begin{equation}
\sum_{a_1,a_2,\dots,a_k}^a\big(\big)\equiv\sum_{a_1+a_2+\dots+a_k=a}\binom{a}{a_1,a_2,\dots,a_k}.
\end{equation}
In this notation the multinomial theorem is represented by
\begin{equation}
(x_1+x_2+\dots+x_k)^a=  \sum_{a_1,a_2,\dots,a_k}^a\big(\big)\prod_{i=1}^k x_i^{a_i},  
\end{equation}
which allows us to express $\omega$ as
\begin{equation}
\omega=\prod_{l=1}^{k}\sum_{\eta^1_{(l)},\dots,\eta^k_{(l)}}^{n_l}\big(\big)\prod_{j=1}^k\left[\sum_{i=1}^{N_j}w_l\left(x_i^j\right)\right]^{\eta^j_{(l)}},
\end{equation}
where the first summation is over all possible combinations of $k$ nonnegative integers $\eta^1_{(l)},\dots,\eta^k_{(l)}$ that sum up to $n_l$. Let us now use the multinomial theorem again, this time to expand the ${\eta^j_{(l)}}$th power:	
\begin{equation}
\omega=\prod_{l=1}^{k}\sum_{\eta^1_{(l)},\dots,\eta^k_{(l)}}^{n_l}\big(\big)\prod_{j=1}^k\sum_{\eta^j_{1(l)},\dots,\eta^j_{N_j(l)}}^{\eta^j_{(l)}}\big(\big)\prod_{i=1}^{N_j}\left[w_l(x_i^j)\right]^{\eta^j_{i(l)}}.
\end{equation}
This formulation could be rearranged to give
\begin{multline}
\omega=\sum_{\eta^1_{(1)},\dots,\eta^k_{(1)}}^{n_1}\big(\big)\sum_{\eta^1_{(2)},\dots,\eta^k_{(2)}}^{n_2}\big(\big)\dots\sum_{\eta^1_{(k)},\dots,\eta^k_{(k)}}^{n_k}\big(\big)\\
\prod_{j=1}^k\sum_{\eta^j_{1(1)},\dots,\eta^j_{N_j(1)}}^{\eta^j_{(1)}}\big(\big)\sum_{\eta^j_{1(2)},\dots,\eta^j_{N_j(2)}}^{\eta^j_{(2)}}\big(\big)\dots\sum_{\eta^j_{1(k)},\dots,\eta^j_{N_j(k)}}^{\eta^j_{(k)}}\big(\big)\\
\prod_{i=1}^{N_j}w_1(x^j_i)^{\eta^j_{i(1)}}\cdot w_2(x^j_i)^{\eta^j_{i(2)}}\cdot\ldots\cdot w_k(x^j_i)^{\eta^j_{i(k)}}.
\end{multline}
If we now put $\omega$ expressed in such way back to equation \ref{e1}, we can notice that integration over $x_i^j$ can be applied  to the product $w_1(x^j_i)^{\eta^j_{i(1)}}\cdot \ldots\cdot w_k(x^j_i)^{\eta^j_{i(k)}}$ to give
\begin{multline}
\label{eq2}
\langle W_1^{n_1}\cdot \ldots\cdot W_k^{n_k} \rangle=\sum_{N_1=0}^\infty\dots\sum_{N_k=0}^\infty \mathrm{P}(N_1,\dots,N_k)\\
\sum_{\eta^1_{(1)},\dots,\eta^k_{(1)}}^{n_1}\big(\big)\sum_{\eta^1_{(2)},\dots,\eta^k_{(2)}}^{n_2}\big(\big)\dots\sum_{\eta^1_{(k)},\dots,\eta^k_{(k)}}^{n_k}\big(\big)\\
\prod_{j=1}^k\sum_{\eta^j_{1(1)},\dots,\eta^j_{N_j(1)}}^{\eta^j_{(1)}}\big(\big)\sum_{\eta^j_{1(2)},\dots,\eta^j_{N_j(2)}}^{\eta^j_{(2)}}\big(\big)\dots\sum_{\eta^j_{1(k)},\dots,\eta^j_{N_j(k)}}^{\eta^j_{(k)}}\big(\big)\\
\prod_{i=1}^{N_j}u_j(\eta^j_{i(1)},\eta^j_{i(2)},\dots,\eta^j_{i(k)}),
\end{multline}
where function $u_j$ is defined as
\begin{multline}
\label{eq:u}
u_j(\eta^j_{i(1)},\eta^j_{i(2)},\dots,\eta^j_{i(k)})\equiv\\
\frac{1}{\langle N_j \rangle}\int w_1(x)^{\eta^j_{i(1)}}\cdot w_2(x)^{\eta^j_{i(2)}}\cdot\ldots\cdot w_k(x)^{\eta^j_{i(k)}}\rho_j(x)\mathrm{d}x.
\end{multline}
Let us now focus on the part of equation \ref{eq2} depending on $j$, which will be denoted by $\lambda^j$,
\begin{multline}
\lambda^j\equiv\sum_{\eta^j_{1(1)},\dots,\eta^j_{N_j(1)}}^{\eta^j_{(1)}}\big(\big)\sum_{\eta^j_{1(2)},\dots,\eta^j_{N_j(2)}}^{\eta^j_{(2)}}\big(\big)\dots\sum_{\eta^j_{1(k)},\dots,\eta^j_{N_j(k)}}^{\eta^j_{(k)}}\big(\big)\\
\prod_{i=1}^{N_j}u_j(\eta^j_{i(1)},\eta^j_{i(2)},\dots,\eta^j_{i(k)}).
\end{multline}
Since each $u_j$ depends on a tuple of values of length $k$, it is convenient to introduce a notation for such tuples:
\begin{multline}
\boldsymbol{\eta}^j_i\equiv(\eta^j_{i(1)},\eta^j_{i(2)},\dots,\eta^j_{i(k)})\\
\boldsymbol{\eta}^j\equiv(\eta^j_{(1)},\eta^j_{(2)},\dots,\eta^j_{(k)})\\
\boldsymbol{\eta}^j_p+\boldsymbol{\eta}^j_q\equiv(\eta^j_{p(1)}+\eta^j_{q(1)},\eta^j_{p(2)}+\eta^j_{q(2)},\dots,\eta^j_{p(k)}+\eta^j_{q(k)}).
\end{multline}
This allows us to express $\lambda^j$ as
\begin{multline}
\label{lambda}
\lambda^i=\\
\sum_{\boldsymbol{\eta}^j_1+\dots+\boldsymbol{\eta}^j_{N_j}=\boldsymbol{\eta}^j}\left[\prod_{l=1}^k\binom{\eta^j_{(l)}}{\eta^j_{1(l)},\dots,\eta^j_{N_j(l)}}\cdot\prod_{i=1}^{N_j}u_j(\boldsymbol{\eta}^j_i)\right],
\end{multline}
where the first summation is over all possible combinations of $N_j$ tuples $\boldsymbol{\eta}^j_1,\dots,\boldsymbol{\eta}^j_{N_j}$ (each containing $k$ nonnegative integers) that sum up to $\boldsymbol{\eta}^j$. We can notice that zero tuples, i.e., $\boldsymbol{\eta}^j_i=(0,0,\dots, 0)$, do not contribute to $\lambda^j$ since $u_j(0,0,\dots,0)=1$. Let us then express the sequence $\boldsymbol{\eta}^j_1,\boldsymbol{\eta}^j_2,\dots,\boldsymbol{\eta}^j_{N_j}$ as a combination of several non-zero tuples.

Let $\Gamma^j$ denote a set of all such combinations possible for $\boldsymbol{\eta}^j$, so that each $\gamma\in\Gamma^j$ consists of $|\gamma|$ different non-zero tuples: $\boldsymbol{\mu}^\gamma_1,\boldsymbol{\mu}^\gamma_2,\dots,\boldsymbol{\mu}^\gamma_{|\gamma|}$, each occurring $m^\gamma_1,m^\gamma_2,\dots,m^\gamma_{|\gamma|}$ times, respectively. There are also $N_j-\sum_{p=1}^{|\gamma|} m^\gamma_p$ tuples equal to zero in the original sequence $\boldsymbol{\eta}^j_1,\boldsymbol{\eta}^j_2,\dots,\boldsymbol{\eta}^j_{N_j}$. We therefore have
\begin{equation}
\forall_{\gamma\in\Gamma^j} \sum_{p=1}^{|\gamma|} m^\gamma_p\cdot\boldsymbol{\mu}^\gamma_p=\boldsymbol{\eta}^j.
\end{equation}
If we use this to compute $\lambda^j$, we get
\begin{multline}
\lambda^j=\sum_{\gamma\in\Gamma^j}\binom{N_j}{m^\gamma_1,m^\gamma_2,\dots,m^\gamma_{|\gamma|},N_j-\sum_{p=1}^{|\gamma|} m^\gamma_p}\\
\Bigg[\prod_{l=1}^k\binom{\eta^j_{(l)}}{m^\gamma_1\ast\mu^\gamma_{1(l)},m^\gamma_2\ast\mu^\gamma_{2(l)},\dots,m^\gamma_{|\gamma|}\ast\mu^\gamma_{{|\gamma|}(l)}}\times\\
\prod_{p=1}^{|\gamma|}[u_j(\boldsymbol{\mu}^j_p)]^{m^\gamma_p}\Bigg]+\mathds{1}[\boldsymbol{\eta}^j=0],
\end{multline}
where $a\ast b$ means that the value $b$ appears $a$ times in the multinomial symbol. The indicator variable is necessary so that in the case of $\boldsymbol{\eta}^j=0$, we have $\lambda^j=1$, as in Eq.~\ref{lambda}.

If we now focus on the multinomial symbol involving $N_j$, it can be expanded as
\begin{multline}
\binom{N_j}{m^\gamma_1,\dots,m^\gamma_{|\gamma|},N_j-\sum_{p=1}^{|\gamma|} m^\gamma_p}=\\
\frac{N_j!}{m^\gamma_1!\cdot m^\gamma_2!\cdot \ldots \cdot m^\gamma_{|\gamma|}!\cdot(N_j-\sum_{p=1}^{|\gamma|} m^\gamma_p)!}=\\
\frac{N_j\cdot(N_j-1)\cdot\ldots\cdot(N_j-(\sum_{p=1}^{|\gamma|} m^\gamma_p-1))}{m^\gamma_1!\cdot m^\gamma_2!\cdot \ldots \cdot m^\gamma_{|\gamma|}!}.
\end{multline}
We can see it is a polynomial of $N_j$ of degree $\sum_{p=1}^{|\gamma|} m^\gamma_p$, which is at least 1 and at most $\eta^j_\Sigma\equiv\sum_{l=1}^k\eta^j_{(l)}$.

Since $\lambda^j$ is a weighted sum of such polynomials (and indicator variable), it is a polynomial of $N_j$ of at most the same degree and can therefore be expressed as
\begin{equation}
\lambda^j=\lambda^j_0+\lambda^j_1 N_j+\lambda^j_2 N_j^2+\dots+\lambda^j_{\eta^j_\Sigma} N_j^{\eta^j_\Sigma}.
\end{equation}
Further coefficients, i.e., $\lambda^j_p$ for $p>\eta^j_\Sigma$, equal zero.

Let us put this formulation of $\lambda^j$ back to Eq.~\ref{eq2}. This gives us
\begin{multline}
\langle W_1^{n_1}\cdot \ldots\cdot W_k^{n_k} \rangle=\sum_{N_1=0}^\infty\dots\sum_{N_k=0}^\infty \mathrm{P}(N_1,\dots,N_k)\\
\sum_{\eta^1_{(1)},\dots,\eta^k_{(1)}}^{n_1}\big(\big)\sum_{\eta^1_{(2)},\dots,\eta^k_{(2)}}^{n_2}\big(\big)\dots\sum_{\eta^1_{(k)},\dots,\eta^k_{(k)}}^{n_k}\big(\big)\\
\prod_{j=1}^k\lambda^j_0+\lambda^j_1 N_j+\lambda^j_2 N_j^2+\dots+\lambda^j_{\eta^j_\Sigma} N_j^{\eta^j_\Sigma}.
\end{multline}
We can rearrange it as
\begin{multline}
\langle W_1^{n_1}\cdot \ldots\cdot W_k^{n_k} \rangle=\sum_{N_1=0}^\infty\dots\sum_{N_k=0}^\infty \mathrm{P}(N_1,\dots,N_k)\\
\sum_{\eta^1_{(1)},\dots,\eta^k_{(1)}}^{n_1}\big(\big)\dots\sum_{\eta^1_{(k)},\dots,\eta^k_{(k)}}^{n_k}\big(\big)\sum_{q_1=0}^{\eta^1_\Sigma}\sum_{q_2=0}^{\eta^2_\Sigma}\dots\sum_{q_k=0}^{\eta^k_\Sigma}\\
\lambda^1_{q_1}\lambda^2_{q_2}\dots\lambda^k_{q_k}\times N_1^{q_1}N_2^{q_2}\dots N_k^{q_k},
\end{multline}
which finally gives us
\begin{multline}
\label{final}
\langle W_1^{n_1}\cdot \ldots\cdot W_k^{n_k} \rangle=\sum_{\eta^1_{(1)},\dots,\eta^k_{(1)}}^{n_1}\big(\big)\dots\sum_{\eta^1_{(k)},\dots,\eta^k_{(k)}}^{n_k}\big(\big)\\
\sum_{q_1=0}^{\eta^1_\Sigma}\sum_{q_2=0}^{\eta^2_\Sigma}\dots\sum_{q_k=0}^{\eta^k_\Sigma}
\lambda^1_{q_1}\lambda^2_{q_2}\dots\lambda^k_{q_k}\langle N_1^{q_1}\cdot N_2^{q_2}\cdot \ldots\cdot N_k^{q_k}\rangle.
\end{multline}

Since $\sum_{j=1}^k\eta^j_\Sigma=n_1+n_2+\dots+n_k=n$, the order of the moments at the right hand side is at most equal $n$. We have therefore just shown how to express any moment of $W$ distributions with order $n$ as a sum of moments of $N$ distributions with order $\leq n$. Since this dependency is linear, we can define the whole problem as a set of linear equations. It will have the coefficients
\begin{equation}
a^{q_1,q_2,\dots,q_k}_{n_1,n_2,\dots,n_k}\equiv\sum_{\eta^1_{(1)},\dots,\eta^k_{(1)}}^{n_1}\big(\big)\dots\sum_{\eta^1_{(k)},\dots,\eta^k_{(k)}}^{n_k}\big(\big)\lambda^1_{q_1}\lambda^2_{q_2}\dots\lambda^k_{q_k},
\end{equation}
where $n_1+n_2+\dots+n_k=n$ and $q_1+q_2+\dots+q_k=n$. We also need to define elements which will take into account the contribution of moments of $N$ with orders lower than $n$:
\begin{multline}
b_{n_1,n_2,\dots,n_k}\equiv\langle W_1^{n_1}\cdot \ldots\cdot W_k^{n_k}\rangle -\sum_{\eta^1_{(1)},\dots,\eta^k_{(1)}}^{n_1}\big(\big)\dots\\ \sum_{\eta^1_{(k)},\dots,\eta^k_{(k)}}^{n_k}\big(\big)\sum_{q_1=0}^{\eta^1_\Sigma}\sum_{q_2=0}^{\eta^2_\Sigma}\dots\sum_{q_k=0}^{\eta^k_\Sigma}\mathds{1}[q_1+q_2+\dots+q_k<n]\\
\lambda^1_{q_1}\lambda^2_{q_2}\dots\lambda^k_{q_k}\langle N_1^{q_1}\cdot N_2^{q_2}\cdot\ldots\cdot N_k^{q_k}\rangle.
\end{multline}

To arrange the moments in a linear order, let us now choose any one-to-one function $f$ from sequences of length $k$ summing up to $n$ to numbers $1, 2, \dots, \binom{n+k-1}{k-1}$. We can use it to construct a matrix $\textbf{A}$ having elements $A_{\xi,\zeta}=a^{q_1,q_2,\dots,q_k}_{n_1,n_2,\dots,n_k}$ and vector $\textbf{B}$ with elements $B_\xi=b_{n_1,n_2,\dots,n_k}$ for $\xi=f(n_1,n_2,\dots,n_k)$ and $\zeta=f(q_1,q_2,\dots,q_k)$. We can also arrange unknown moments in a vector $\textbf{N}$ such that $N_\zeta=\langle N_1^{q_1}\cdot N_2^{q_2}\cdot\ldots\cdot N_k^{q_k}\rangle$. This allows us to express equation \ref{final} as
\begin{equation}
\sum_{\zeta}A_{\xi,\zeta}N_\zeta=B_\xi,
\end{equation}
or in matrix notation:
\begin{equation}
\textbf{A}\textbf{N}=\textbf{B}.
\end{equation}
If $\text{det} \textbf{A}\neq 0$, the moment we are looking for can be computed as
\begin{equation}
\textbf{N}=\textbf{A}^{-1}\textbf{B}.
\end{equation}

\section{Modifications}
\label{sec_modifications}
In the previous section we have shown the procedure to compute the $n$th moments of the multiplicity distribution as a generalisation of the computations for $n=2$ and $n=3$, but to apply it in practice we needed to make three modifications.

Firstly, to compute the first moments as proposed in Ref.~\cite{Rustamov:2012bx}, we need to replace Eq.~\ref{eq:density} with the following:
\begin{equation}
\int \mathrm{d}x \rho_j(x)=\langle A_j \rangle.
\end{equation}
Now the distribution of a measured $x$ for a given particle type $j$ is normalised to arbitrary value $A_j$, which does not have to equal $N_j$. As a result, we also need modify Eq.~\ref{eq:u}, which now becomes
\begin{multline}
u_j(\eta^j_{i(1)},\eta^j_{i(2)},\dots,\eta^j_{i(k)})\equiv\\
\frac{1}{\langle A_j \rangle}\int w_1(x)^{\eta^j_{i(1)}}\cdot w_2(x)^{\eta^j_{i(2)}}\cdot\ldots\cdot w_k(x)^{\eta^j_{i(k)}}\rho_j(x)\mathrm{d}x.
\end{multline}
The rest of the procedure holds, and corrected $\langle N_j\rangle$ could be computed by applying it for $n=1$.

Secondly, the measured $x$ is traditionally associated with the particle mass, but it can be any measured quantity, not necessarily a single scalar value. In general, it could be a multi-dimensional vector $\mathbf{x}$, e.g. mean energy loss and time-of-flight, as long as integration in function $u$ is performed accordingly.

Thirdly, measurement of $x$ could be performed in several phase space bins, corresponding to different detector configurations. In such cases Eq.~\ref{eq:density} takes form
\begin{equation}
 \sum_{\theta\in\Theta}\int\mathrm{d}x \rho_j(x,\theta)=\langle N_j \rangle,
\end{equation}
where $\theta$ denotes a configuration from a configuration space $\Theta$. Analogously, the definition of $w$ (equation \ref{eq:w}) has to take into account $\theta$ as well:
\begin{equation}
w_j(x,\theta)\equiv\frac{\rho_j(x,\theta)}{\sum_{l=1}^k \rho_l(x,\theta)},
\end{equation}
where $w_j(x,\theta)$ denotes value of the $j$th identity variable for a measurement $x$ registered in configuration $\theta$. Finally, the computation of $u$ (equation \ref{eq:u}) has to take into account measurements in all configurations, so
\begin{multline}
u_j(\eta^j_{i(1)},\eta^j_{i(2)},\dots,\eta^j_{i(k)})\equiv\frac{1}{\langle N_j \rangle}\sum_{\theta\in\Theta}\\
\int w_1(x,\theta)^{\eta^j_{i(1)}}\cdot \ldots\cdot w_k(x,\theta)^{\eta^j_{i(k)}}\rho_j(x,\theta)\mathrm{d}x,
\end{multline}

All three modifications have been described here separately for simplicity, but could be combined if necessary.

\section{Implementation}
\label{sec_idhim}
The {\it Idhim} program was designed to provide an easy way to obtain moments of the true multiplicity distribution of identified particles provided the detector resolution is know. 

The implementation in \textit{Java}, using \textit{EJML}\footnote{\url{http://ejml.org}} library for linear algebra operations, is available as open source\footnote{\url{https://github.com/piotrmp/idhim}}. The required input to the program includes:
\begin{enumerate}[(i)]
    \item a list of particle types in a text file, with each line providing a particle type name,\label{types}
    \item $\langle W_{1}^{n_1}\cdot\ldots\cdot W_k^{n_k}\rangle$ moments in a tsv (i.e., tab-separated values) file, with each line describing one moment as a list if $n_1,\ldots,n_k$ indices, followed by the moment value,\label{means}
    \item a list of phase space bins, where a detector response is known, as a tsv file (if there is more than one kinematic variables which define such bins, multiple tab-separated indices may be provided),\label{bins}
    \item a directory containing files with a detector response functions in each bin.\label{rhos}
\end{enumerate}
An exemplary set of all needed files is provided with the program.

The input format allows for applicability to a wide range of different experiments. Firstly, a number of considered particle types is arbitrary. In a typical case of particle identification it depends on a collision energy and available statistics, e.g., at low interaction energies one does not need to consider deuterons and/or Helium-3, whereas at high energies or with large available statistics they must be taken into account. Secondly, only $\langle W_1^{n_1}\cdot \ldots\cdot W_k^{n_k} \rangle$ moments, not the full distributions, need to be provided.  
\begin{figure}
\centering
  \includegraphics[width=0.3\textwidth]{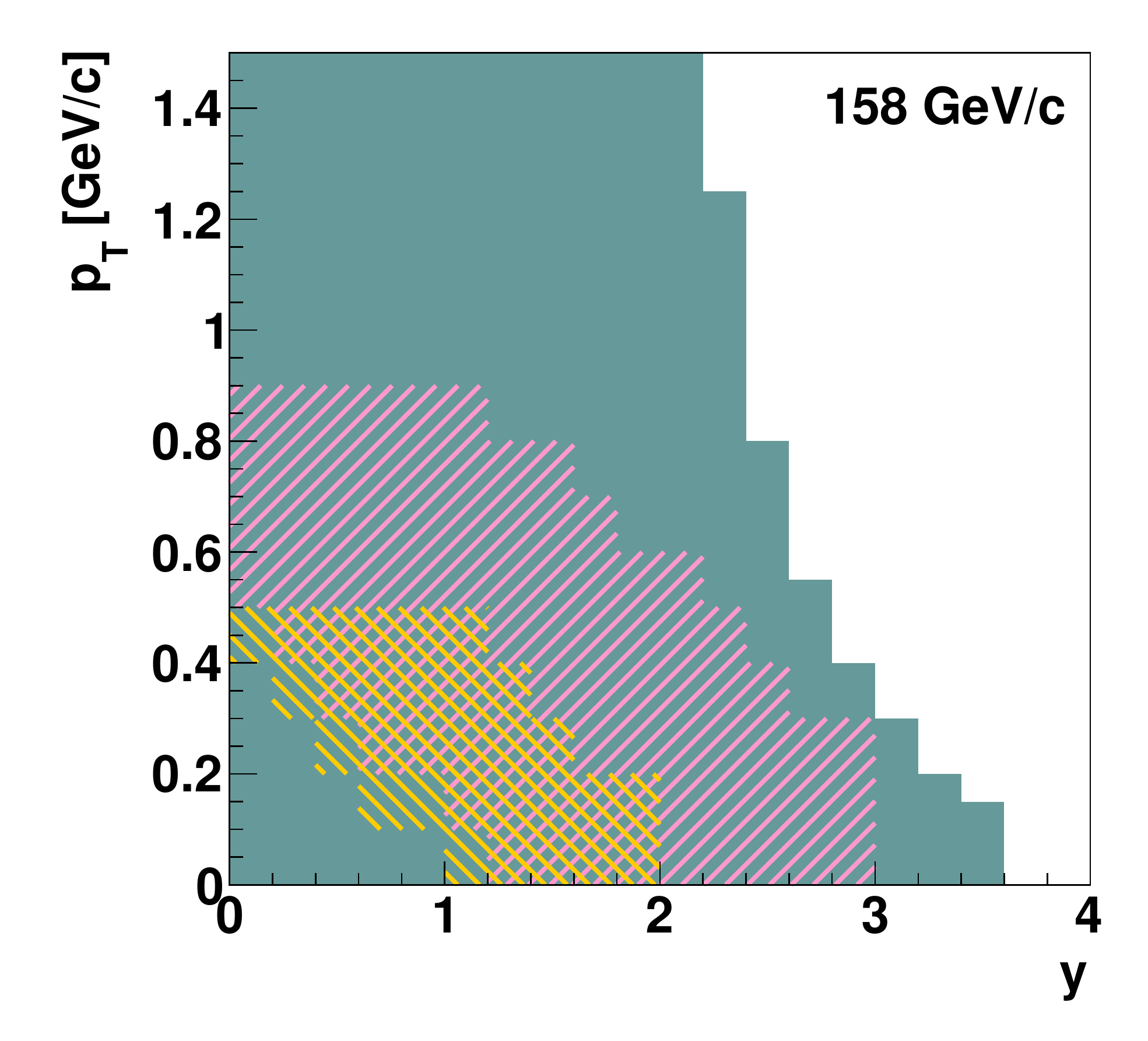}
\caption{NA61/SHINE detector acceptance (solid area) with indicated region where particles are identified via their energy loss (magenta stripes) and their time-of-flight (yellow stripes) in p+p interactions at $\sqrt{s_{NN}}=17.3$ GeV.}
\label{fig:acc}       
\end{figure} 
{Finally, in a typical experiment, a particle identification is performed by a set of detectors with an overlapping momentum coverage. Thus, a full momentum coverage of an experiment consists of regions with $\rho$ being 1D function, e.g., when particles are identified only by $dE/dx$ or time-of-flight (ToF), or 2D function, e.g., when particles are identified by combined measurements of $dE/dx$ and ToF. An example of such a non-uniform detector acceptance is shown in Fig.~\ref{fig:acc}. Bins with any number of dimensions, which reflect changing detector configuration or particle yields, can be defined as long as density function for all particle types is given in the same points of the space. 
} 

The next section includes example demonstrating the usefulness of the program features described above. 

\section{Test on simulated data}
\label{sec_tests}
The computation of all moments of multiplicity distributions up to the fourth order was tested on two models. The first one is a Monte Carlo model (so-called fast generator), where the number of particles of a given type produced in a single event was generated from Poisson distributions with a different free parameter $\lambda$ for each considered particle type. Test included four most popular particle types, namely electrons, pions, kaons and protons, with their respective $\lambda$ as 1, 10, 2, 4. The number of events is set to 1,000,000. 

Particles are generated according to the Poisson distribution and are uncorrelated (except the detector response), so the true values of generated moments are
\begin{align}
\langle N_{j} \rangle = \lambda,\\
\langle N_{j}^{2}\rangle = \lambda(1+ \lambda),\\
\langle N_{j}^{3}\rangle = \lambda(1+ 3\lambda + \lambda^{2}),\\
\langle N_{j}^{4}\rangle = \lambda(1+ 7\lambda+6\lambda^{2}+\lambda^{3}).
\end{align}
The generated cross-moments are defined as the multiplication of the pure ones. 
\begin{figure}
  \includegraphics[width=0.45\textwidth]{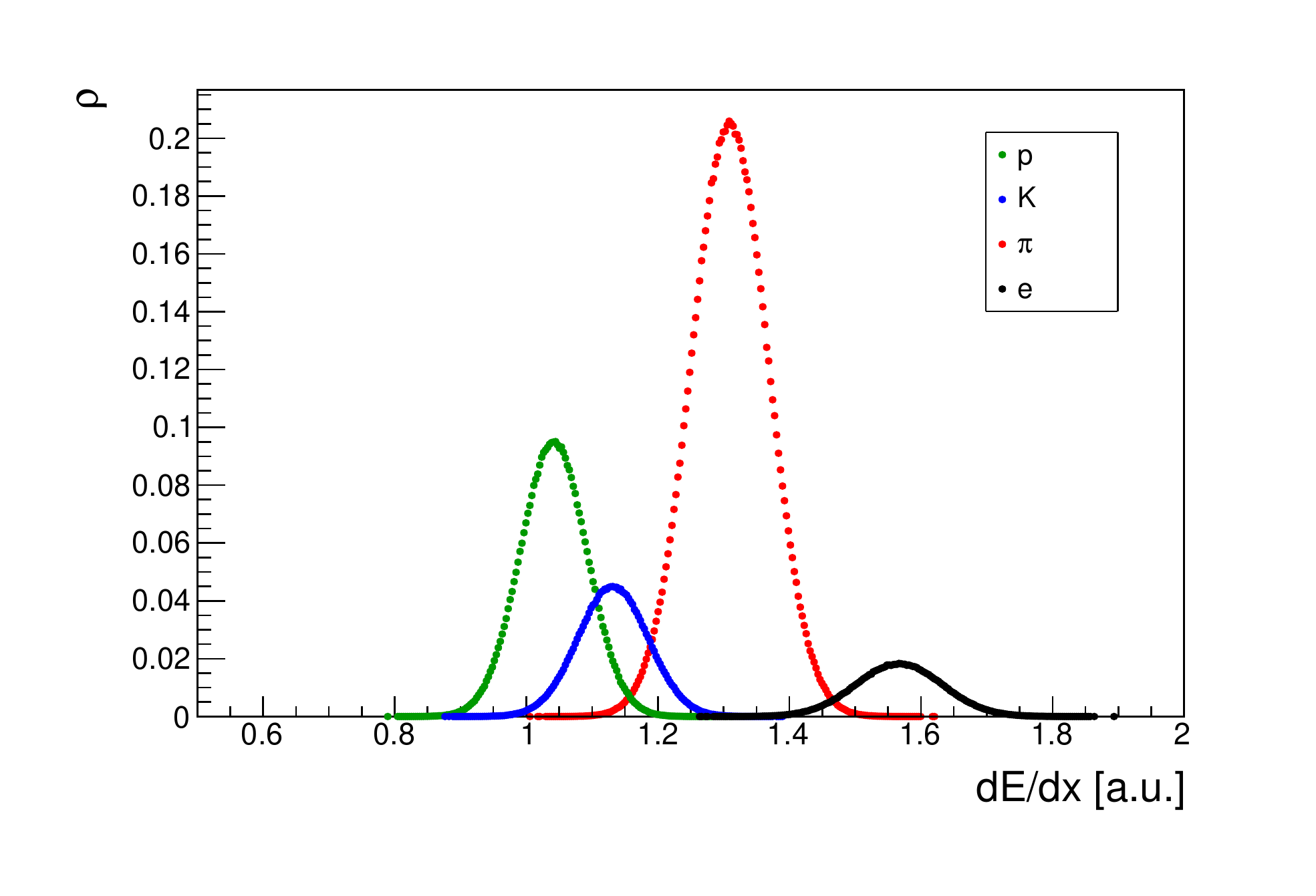}
\caption{Generated $dE/dx$ distribution in a single bin. For details see text.}
\label{fig:1}       
\end{figure}

A simulated detector response consists of mean energy loss measurements in the Time Projection Chamber. For each particle, its mean energy loss was generated from a Gaussian distribution with parameters based on experimental data from Refs.~\cite{Marco_fit2,Aduszkiewicz:2017sei} in two bins simulating the momentum dependence of the detector response. Testing several different momentum dependencies showed that particle distribution between bins does not affect the final results. An exemplary simulated dE/dx distribution in a single bin is shown in Fig.~\ref{fig:1}.

The {\it Idhim} program is used to obtain reconstructed moments of the considered particle types up to the fourth order. The statistical uncertainty of the reconstructed moments results from uncertainty of the fitted distributions $\rho_{j}(x)$ as well as from the $\langle W_{1}^{n_1}\cdot\ldots\cdot W_k^{n_k}\rangle$ moment values. Both sources are correlated, so the standard error propagation is complicated and inconvenient. Instead, the statistical uncertainty is obtained using the bootstrap method~\cite{efron}.

\begin{figure*}
\centering
  \includegraphics[width=0.9\textwidth]{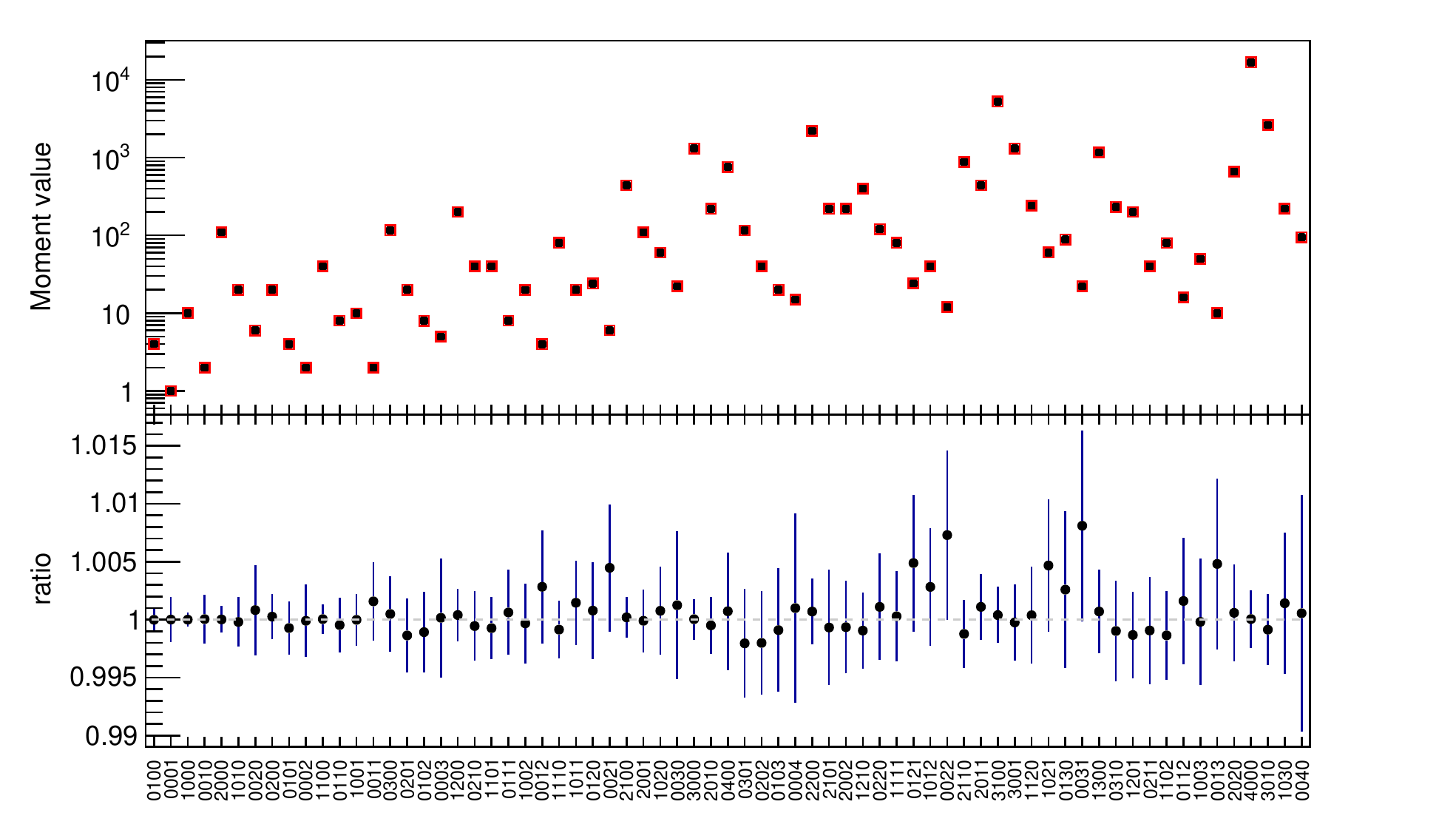}
\caption{ Reconstructed (black circles) and generated (red squares) moments and their ratio. 
}
\label{fig:2}       
\end{figure*}
Reconstructed and generated moments as well as their ratio are shown in Fig.~\ref{fig:2}. The ratio is 1 within the statistical uncertainty for all considered values.
\begin{figure}
\centering
  \includegraphics[width=0.45\textwidth]{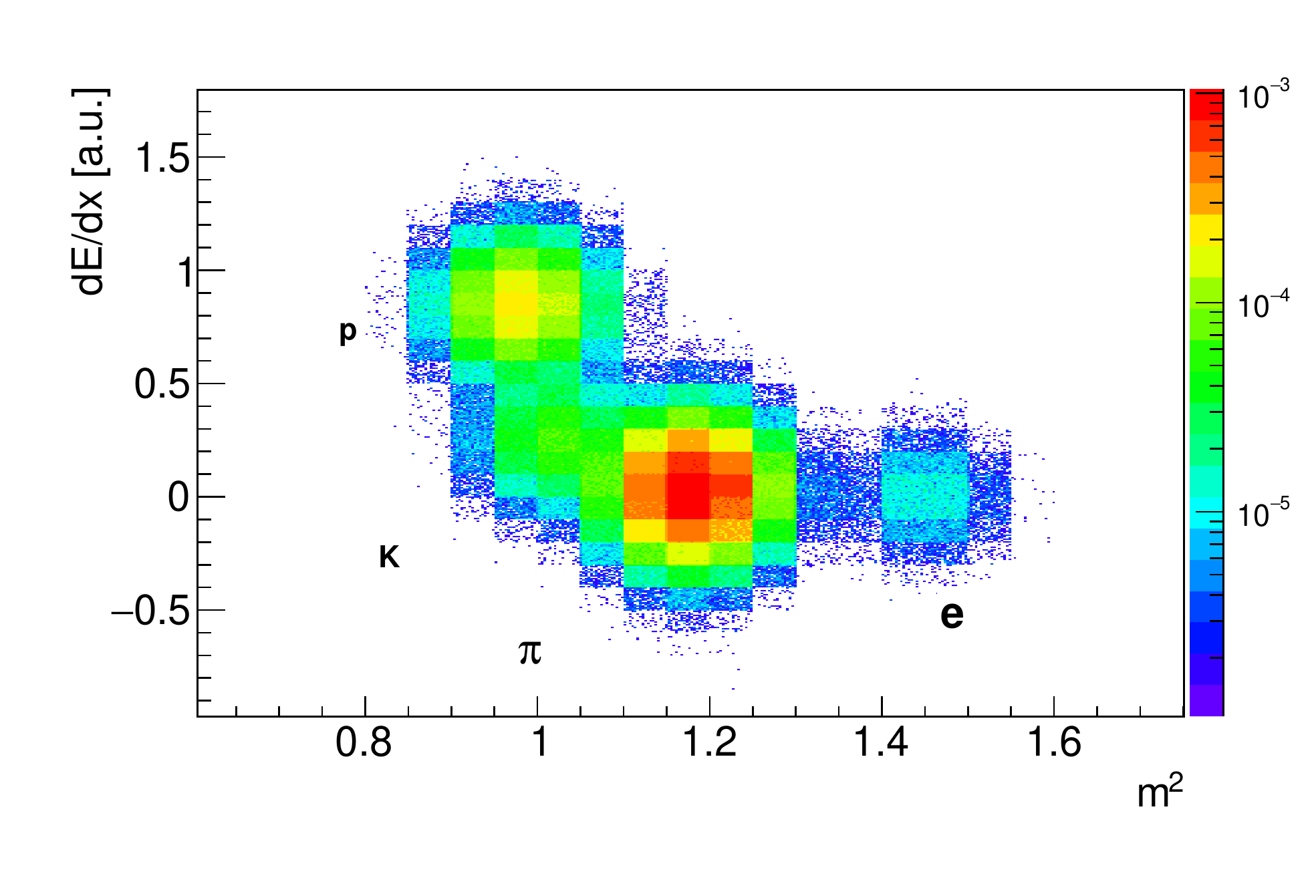}
\caption{Generated dE/dx and $m^{2}$ distribution in p+p interactions at 158 GeV/c beam momentum in EPOS.}
\label{fig:3}       
\end{figure}

\begin{figure*}
\centering
  \includegraphics[width=0.9\textwidth]{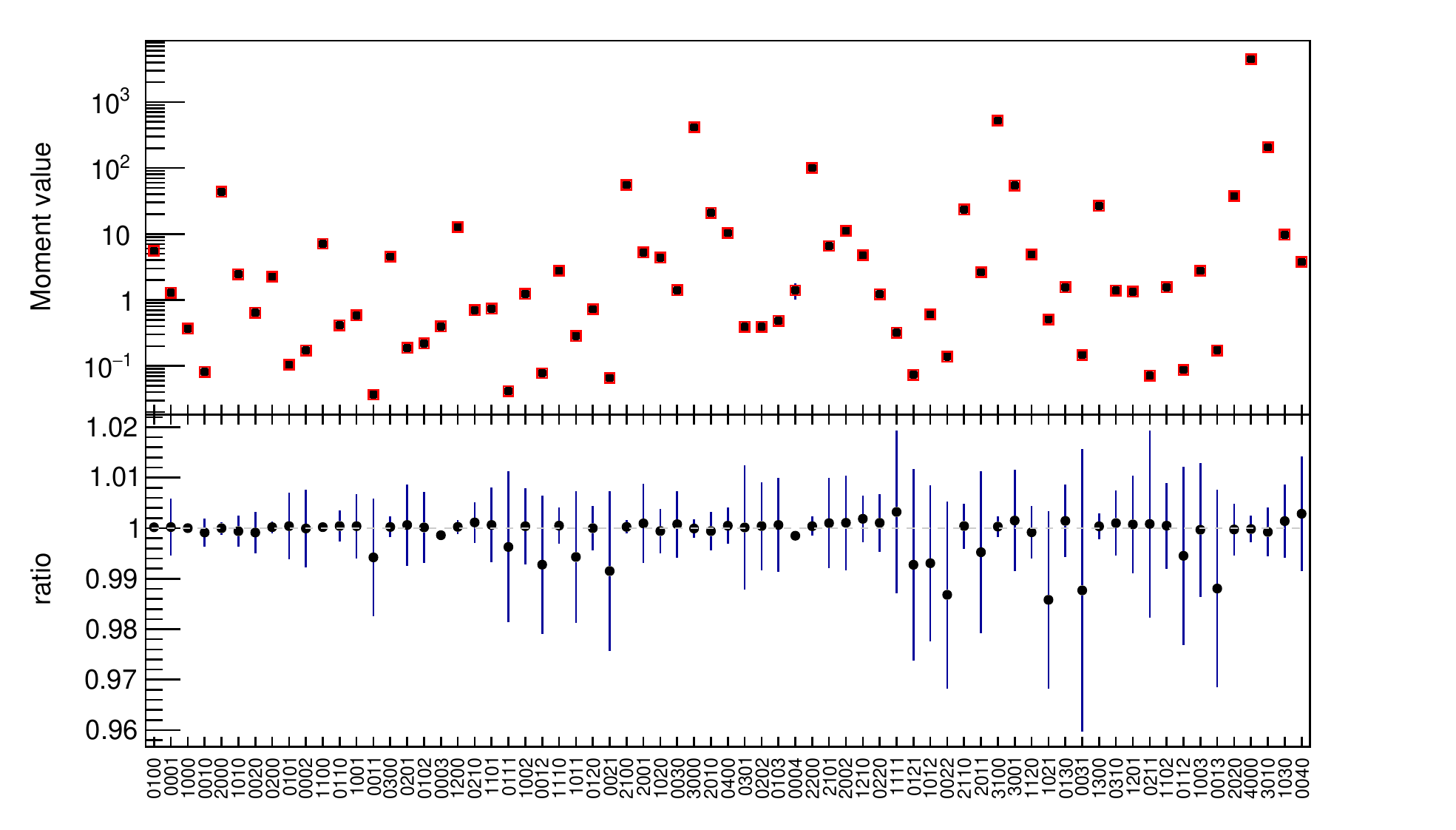}
\caption{Reconstructed (black circles) and generated (red squares) moments and their ratio. The ratio uncertainty values for moments 0003 and 0004, omitted from the plot for clarity, equal 1.00$\pm$0.07 and 1.00$\pm$0.55, respectively
}
\label{fig:4}       
\end{figure*}
Another test was performed using 3 million p+p interactions at $\sqrt{s_{NN}}=17.3$ GeV generated using the EPOS~\cite{Werner:2008zza,Pierog:2009zt} model with the detector acceptance containing two types of acceptance regions --- $dE/dx$ only and combined ToF and $dE/dx$. An example of such a two dimensional distribution is shown in Fig.~\ref{fig:3}. The shape of the 2D distribution and its parameters were based on a real data analysis in Ref.~\cite{Kuich:2017mlh}. Again, in order to mimic the momentum dependence of the detector response, it was divided into several bins.

Reconstructed and generated moments as well as their ratio are shown in Fig.~\ref{fig:4}. Again, the ratio is 1 within the statistical uncertainty for all considered values.

Both the procedure and its implementation are functioning as expected. The difference between generated and reconstructed first moments of $N$ and $W$ is negligible but in case of the higher orders, the differences can reach 70$\%$. In order to accommodate for different possible shapes of the $\rho$ functions, they are delivered in a binned form. Thus, a proper binning is important to describe the functions' shapes. The identity method does not address other detector biases or its efficiency. Other possible biases should be addressed by the appropriate experimental tools (for examples and details see Refs.~\cite{Aduszkiewicz::2015jna,Bzdak:2016qdc,LorenzCPOD}).   

\section{Conclusion}
\label{sec_conclusion}
{ In this paper we extend the identity method in two ways. Firstly, a new strict procedure to obtain $n$th order moments of multiplicity distribution of an arbitrary number of particles is discussed. Secondly, a software implementation of this procedure is presented.}
Provided a detector response is known, it computes any moments, including the first ones. It is equally precise both for low and high mean multiplicities. Two tests were performed in order to validate the program. The first test, based on simple fast generator check, showed that program works well in case of lack of correlations between particles. The difference between the reconstructed and generated moments is at the level of statistical uncertainty or below. The second test was performed on p+p interactions simulated in the EPOS model. The second test confirmed that correlations between particles do not affect the program's efficiency. It also showed that {\it Idhim} can be easily used in the case of a non-uniform detector acceptance which contains different detector types.

As a last comment we would like to stress that the successful analysis of moments of identified particle distributions depends on an understanding of a detector response. Possible flaws in description of the $\rho$ functions will propagate to the identity method and the final results. Moreover, the identity method does not compensate for a limited detector efficiency. Thus, $\rho$ distributions and mean $\langle W \rangle$'s have to be corrected for a limited and often momentum-dependent detector efficiency by other known methods.

\begin{acknowledgements}
We would like to thank M. Gazdzicki for fruitful discussions and comments.
The  work  of M.M.P. was  partially  supported  by  the  National  Science
Center, Poland grant 2015/18/M/ST2/00125.
\end{acknowledgements}

\end{document}